\documentclass[lettersize,journal]{IEEEtran}
\usepackage{amsmath,amsfonts}
\usepackage{algorithmic}
\usepackage{algorithm}
\usepackage{array}
\usepackage[caption=false,font=normalsize,labelfont=sf,textfont=sf]{subfig}
\usepackage{textcomp}
\usepackage{stfloats}
\usepackage{url}
\usepackage{upgreek}
\usepackage{verbatim}
\usepackage{graphicx}
\usepackage{cite}

\begin{document}

\title{Computationally Efficient Nanophotonic Design through Data-Driven Eigenmode Expansion}

\author{Mehmet Can Oktay and Emir Salih Magden* \\
Department of Electrical and Electronics Engineering, Koç University, Sarıyer, 34450, Istanbul, Turkey
\thanks{*corresponding author: esmagden@ku.edu.tr}}

\maketitle

\begin{abstract}
Growing diversity and complexity of on-chip photonic applications requires rapid design of components with state-of-the-art operation metrics. Here, we demonstrate a highly flexible and efficient method for designing several classes of compact and low-loss integrated optical devices. By leveraging a data-driven approach, we represent devices in the form of cascaded eigenmode scattering matrices, through a data-driven eigenmode expansion method. We perform electromagnetic computations using parallel data processing techniques, demonstrating simulation of individual device responses in tens of milliseconds with physical accuracies matching 3D-FDTD. We then couple these simulations with nonlinear optimization algorithms to design silicon-based waveguide tapers, power splitters, and waveguide crossings with state-of-the-art performance and near-lossless operation. These three sets of devices highlight the broad computational efficiency of the design methodology shown, and the applicability of the demonstrated data-driven eigenmode expansion approach to a wide set of photonic design problems. 
\end{abstract}


\section{Introduction}
\IEEEPARstart{D}{evelopments} in integrated photonics and the extending application space continue to drive the need for compact, highly efficient building blocks for next-generation applications [1,2]. With stringent application requirements, performance metrics of individual photonic building blocks, including low insertion losses and wide operation bandwidths [3-5], are more important than ever to ensure reliable and consistent optical system performance in applications ranging from optical communications and interferometry to on-chip sensing and signal processing [6,7]. Designing high-performance optical structures using only physical intuition and knowledge of fundamental wave coupling mechanisms can be prohibitive due to the limited degrees of freedom available in device geometry for some applications [8]. 

In recent years, machine learning-based design methods have become popular for designing compact and high-performance photonic devices [9]. In these approaches, a design space consisting of geometrical parameters of the device is sampled and searched for the most suitable optical response [10,11]. These approaches are typically coupled with FDFD [12] or FDTD [13] simulations that estimate individual devices' performance during this optimization process. While these specific electromagnetic simulations are known to be physically accurate, their computationally demanding nature results in design tasks that can be computationally prohibitive, especially when a large number of iterations are required in order to achieve the required device performance. On the other hand, the use of eigenmode expansion (EME) as a computational tool for the inverse design of photonic components remains largely unexplored, Traditional EME, despite being fully vectorial and 3D-accurate, is mainly used in larger photonic devices, as it must first section a given device geometry into consecutive sections and computes the optical scattering matrices through. However, when used in conjunction with modern data processing techniques, EME can be highly efficient to implement for iterative device simulations. This avenue presents tremendous opportunities in the design of compact, broadband, and ultra-high-performance photonic devices for a large class of on-chip optical applications.

In this paper, we leverage the synergy between machine learning and the eigenmode expansion method to create a new data-driven approach to photonic device simulation and optimization. We bypass the computational bottlenecks inherent in traditional simulation methods by creating a comprehensive database of waveguide parameters over a wide range of widths and wavelengths. Using this database and parallel computational mechanisms, we demonstrate the simulation of individual devices in several tens of milliseconds with physical accuracy matching that of 3D-FDTD, instead of typical timeframes of minutes or hours required for other traditional electromagnetic simulations. Together with the nonlinear optimization mechanisms, this simulation approach enables the rapid calculation and optimization of device responses with flexible design parameters, simultaneously enabling wide functional capability and computational efficiency in photonic device design. We showcase this design paradigm in a silicon-based platform through a series of demonstrations, including designs of high-performance and low-loss integrated tapers, 1x2 power splitters, and waveguide crossings. With computational speedups exceeding five orders of magnitude over conventional methods, this approach also opens up new avenues for designing application-specific, high-performance, and broadband-integrated optical systems.

\section{Data-Driven Eigenmode Expansion Method}
The eigenmode expansion (EME) method is based on calculating the mode-specific amplitude scattering coefficients between eigenmodes through consecutive sections of a given device geometry. In this method, the optical scattering matrices through each section and the mode overlaps at interfaces between the adjacent sections are computed and then sequentially cascaded to obtain the device's overall performance. EME also allows for the simulation of a given device at many different lengths with little added computational load, which is especially important for the nanophotonic design of slowly varying structures. Typical use of the EME method involves computing scattering matrices of devices with predetermined transverse structures, and estimating their transmission as a function of device length to minimize excitation of unwanted modes at the output. As a fully vectorial and bidirectional method, EME provides complete solutions to Maxwell’s equations and can accurately simulate many classes of devices [14]. However, an important drawback is that the eigenmode computations and their overlaps are specific for the exact transverse device geometry and material platform chosen, including the core and cladding materials, layer heights, and waveguide widths. As a result, simulating a brand-new device design with EME requires recalculating eigenmodes and their overlaps at the newly determined transverse slices and interfaces. These computationally extensive operations preclude the use of traditional EME in machine learning-based, iterative optimization methods, which require repetitive simulations of device performance throughout many iterations as design parameters are updated.  

To develop a machine learning-compatible EME method, we first develop an efficient and rapid algorithm for calculating the optical scattering matrix for a given nanophotonic device. Due to its widespread availability, we focus primarily on the 220 nm-thick silicon-on-insulator (SOI) platform and restrict our set of devices to those made from a strip waveguide with a continuously varying width in the propagation direction. This allows us to represent a large class of devices, including various tapers, splitters, or crossings, using only an array of waveguide widths, W[i], as shown in Fig. 1(a). This array of widths are used to create the continuous waveguide geometry using a cubic spline interpolation [15] in the propagation direction. While we used cubic splines for all demonstrations in this paper, other interpolation methods resulting in continuous geometries may also be used for the device representation.

To accurately calculate the optical scattering matrix resulting from these specific eigenmodes, we first uniformly sample the cubic spline in the z-direction, yielding a list of widths for the entire device. Each one of these uniformly spaced samples is then rounded to the nearest existing width in a set of predetermined widths that are stored in a local database. The initial uniform sampling is iteratively repeated with increasing resolution, until no further changes in the final geometry are observed. This operation ensures that the device is represented using only the widths existing in this database, in a piecewise constant fashion. A representative example is shown by the solid black curve in Fig. 1(a). As an inherent benefit of this data-driven approach, the resulting transverse slices are nonuniformly spaced in the propagation direction. Through this conformal representation, regions of the device with greater changes in width are discretized finer (using more frequent slices in the z-direction); and regions with smaller changes in width are discretized coarser. This adaptive approach can more accurately represent a given geometry with an equal or fewer number of slices than conventional uniform sampling, thereby improving computational performance in simulation.

Our database is illustrated in Fig. 1(b), and consists of eigenmode simulation results for the first 25 modes of 300 logarithmically spaced waveguide widths from 400 nm to 12 $\upmu$m. These results include the effective indices of each mode as well as the results of overlap integrals between pairs of all spatial mode distributions of waveguides with consecutive widths. Here, we choose a logarithmic spacing between consecutive widths in this database, to accurately capture the faster change in effective indices and spatial mode distributions at narrower widths. Similar results can also be demonstrated with databases with linearly-spaced widths, at the expense of storing and processing data at significantly more waveguide widths, which directly affect computational performance. It may also be possible to use fewer widths, if the database is only being constructed for more specific devices like compact splitters. However, we note that the conformal device representation in Fig. 1(a) is independent of the database construction, as other spacings (linear, polynomial, etc.) of prerecorded widths would also result in a similar adaptive slicing of the device in the propagation direction. The overlap results are used to model scattering physics at positions of slices determined through the discretization procedure above. Here, the number of tracked modes effectively determines the resulting spatial resolution of the simulation [14]. However, a greater number of modes also results in larger scattering matrices, negatively affecting simulation performance. Considering this tradeoff, we limit the number of modes to 25 for balancing physical accuracy with computational efficiency. For all eigenmode and overlap calculations, a maximum spatial grid resolution of 10 nm was used. This database was created using a single eigenmode expansion simulation, where the effective indices, eigenmodes and mode overlaps are calculated at the boundaries between the predetermined widths in Fig. 1(b). The computational requirements for this process are exactly the same as those required for a typical eigenmode expansion simulation, about 100 minutes on a current desktop computer.

The individual scattering matrices in Fig. 1(b) are shown in red. These are 50x50 matrices created from the overlap integrals, representing mode-by-mode forward and backward scattering coefficients between all 25 modes calculated. The vectors shown in blue are used to depict the effective indices of these 25 modes. While the scattering matrices are calculated for consecutive widths, the effective indices are calculated at each waveguide width directly. Accessing this database allows us to rapidly retrieve all the necessary physical parameters that collectively describe the propagation of modes in a given geometry, without having to recalculate any eigenmodes or overlaps at the time of the simulation. This procedure is illustrated in Fig. 1(c). For a given device geometry, the necessary propagation matrices are created from the recorded effective indices; and the relevant scattering matrices are directly read from the database. These matrices are then multiplied using the Redheffer Star Product (see Appendix) to compute the optical scattering matrix of the entire device [16].

Importantly, as the star product is associative, the products of adjacent scattering matrices can be computed simultaneously, allowing us to take advantage of asynchronous and parallel computational mechanisms enabled by modern graphical processing units (GPUs). Even though the total number of required operations remains the same, the overall computation is performed much faster in this parallel manner, in log2(N) steps to calculate the star product of N individual scattering matrices. We also note that despite the computational efficiency of the method, the resulting optical simulation remains 3-dimensional and physically accurate due to the high spatial resolution and the fully vectorial nature of underlying eigenmodes and overlaps.

While most of the specific operations necessary for each star product are typical matrix products (see Appendix), the calculation of inverses $[I-S_{11}^{B}S_{22}^{A}]^{-1}$ and $[I-S_{22}^{A}S_{11}^{B}]^{-1}$ are much more computationally intensive. However, despite the general complexity of calculating matrix inverses, we note that the blocks $S_{11}^{B}$ and $S_{22}^{A}$ (and hence also their product) are nearly zero in magnitude, as there is minimal mode mismatch or reflection at interfaces between waveguides of consecutive widths in our database. Consequently, we use the Neumann approximation [17] to estimate these inverses as $(I-S)^{-1} \approx \sum_{k=0}^{K} S^{k}$, where K denotes the expansion order. In our calculations, we evaluate this approximation using four terms (K=3), after which we see only negligible change in the convergence of the result.

\begin{figure*}
\centering
\includegraphics[width=7in]{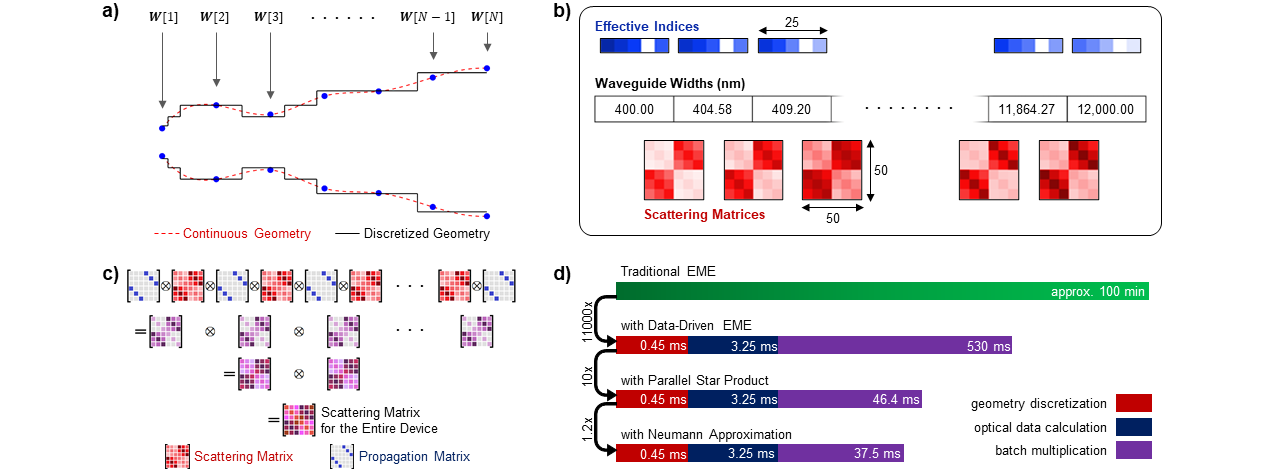}%
\label{fig_first_case}
\caption{a) Representative interpolation of widths and the discrete representation of device geometry. b) Database including predetermined widths, effective indices, and scattering matrices. c) Calculation of device response using the star product. d) Comparison of computational performance from 100-min-long traditional EME to 41.2-ms-long data-driven EME, representing a speedup of 5 orders of magnitude.}
\label{fig_1}
\end{figure*}

Using this data-driven approach together with the computational methods outlined, our eigenmode expansion implementation can calculate the optical scattering matrix of a given device many orders of magnitude times faster than conventional EME, as well as other well-known simulations such as finite-difference time-domain or frequency-domain methods. This is illustrated by the comparative chart in Fig. 1(d) detailing the computational timeframe required for the simulation of an example device, including interpolation and sampling of waveguide widths, eigenmode and overlap calculations or their retrieval from our database, and the necessary star product operations. The comparison starts with an example device simulated using 300 EME cells in Ansys-Lumerical Mode Solutions [18] within a 15$\upmu$m×3$\upmu$m cross-section. This simulation is equivalent to creating our database, and takes approximately 100 minutes using a modern desktop computer. Replacing eigenmode solution and overlap calculations with database retrieval of these parameters, we reduce the device simulation time to about 533 ms, enabling a speedup of over 11000x. The additional computational advantage through GPU implementation of parallel star products and the matrix inverse approximation enables another 10x speedup, with resulting device simulation times as short as 41.2 ms. With this approach, the simulation of a single device is completed over 5 orders of magnitude faster than a traditional EME simulation. Here, we note that a single, common database is sufficient for most device simulations with continuous waveguide geometries. The same database is reused for simulating the same device with modified geometrical parameters or for simulating entirely different devices. Consequently, while creating the database requires the same computational effort as a single traditional EME simulation, this process is performed independently and is not a direct part of the device simulation times reported. All device simulations were performed using a single core on an Intel 9700 CPU and a Tesla V100 GPU through Google Colaboratory Cloud service.

Performing rapid and physically accurate simulations of the device response has important implications for designing many high-performance integrated photonic components. These advantages enabling the rapid simulation of photonic devices present previously elusive opportunities for optimizations where many simulations are necessary as the device geometry is iteratively updated. Coupled with state-of-the-art optimization methods, this technique can enable the efficient design of compact, broadband, and high-performance photonic components for many applications. In the following sections, we demonstrate this capability by designing and optimizing several nanophotonic devices, including a waveguide taper, a power splitter, and a waveguide crossing, all using the rapid eigenmode expansion method we outlined above.

\section{Design of a Waveguide Taper}
We first demonstrate the device optimization capabilities of Rapid Optical Eigenmode Optimizer (ROMEO) outlined above through the design of a waveguide taper. Such tapers are commonly used as transition components as well as spot-size converters for nanophotonic edge and grating couplers. For this specific device, we choose a total of 30 optimizable widths as denoted by W[i] (i=1 through 30) in Fig. 2(a). The taper itself is laterally symmetric and is represented by a cubic spline interpolation of these widths through a device length of 30 $\upmu$m. The input and output waveguide widths for this specific taper are selected to be 500 nm and 9 $\upmu$m, respectively; but these widths can be arbitrarily specified.

Our device design procedure evaluates the entire scattering matrix for this taper as a function of W[i], resulting in the complex mode-to-mode amplitude coefficients for all eigenmodes. We then configure a nonlinear optimizer to maximize the objective function defined as the squared magnitude of the optical transmission coefficient between fundamental transverse electric (TE) modes at the input and output planes. The initial guess for the device is specified with linearly increasing widths in W[i], starting from the user-specified input waveguide and extending to the output waveguide, as this represents a physically motivated starting point. The device is optimized as W[i] are iteratively updated using an open-source implementation [19] of the constrained version of local, non-linear, gradient-free COBYLA algorithm [20]. In contrast to fixed analytical or adiabatic representations of tapers [21], this procedure allows the optimizer to modify device geometry nonlinearly throughout the design process.

Through 277 iterations of this optimization process, the power coupled to the fundamental TE mode at the output of the device is maximized to 99.0\% (-0.0436dB) at the wavelength of $\lambda$=1550 nm, within a total timeframe of 18 seconds. The results in Fig. 2(b) demonstrate the progress of the device objective and constraints during this optimization process. The first plot illustrates the gradual optimization of the device objective (transmission to the fundamental TE mode at the output plane), towards the 0dB transmission goal. Throughout the iterations, the optimization algorithm evaluates possible update directions and magnitudes, by performing multiple different device simulations while gradually modifying the widths W[i]. The specific path taken through the objective function's topology depends on the optimization algorithm chosen [19]. The transmission objective experiences numerous drops and fluctuations during optimization, as the algorithm searches for the best path to the target transmission by exploring different device geometries. After 277 iterations, ROMEO arrives at the optimal solution where a relative change of less than $10^{-5}$ is accepted as the convergence condition. During the optimization process, it is also possible to track the power coupled to any one of the higher order modes. Minimizing the total power coupled to these modes is equivalent to maximizing the power coupled to the fundamental TE mode at the output of the device, and yields the same optimization result. 

In addition to maximizing the optical transmission to the desired optical output mode, we also impose several constraints on the waveguide geometry. To prevent abrupt changes in the waveguide width and help minimize the excitation of higher-order modes, we include a first derivative constraint where $dw(z)⁄dz<2$ for the entire taper. We also impose an additional requirement on this width's curvature ($\kappa$) and constrain the optimizer so that $\kappa < 4 um^{-1}$. While optimizing the transmission to the target output mode, the optimizer also ensures that the final device meets the specified constraints. In the second part of Fig. 2(b), we plot these derivative and curvature constraints with the red and blue curves, respectively. The results demonstrate that the optimization algorithm may temporarily explore geometries that violate these constraints to potentially find better solutions in subsequent iterations. Ultimately, both constraints remain well below their thresholds (dashed lines) at the end of optimization, indicating a fabrication-compatible and smooth taper structure.

In order to verify the performance of this taper, we perform 3D-FDTD simulations with a TE-mode input, using the Ansys Lumerical software suite [18]. We plot the resulting profile of the electric field intensity in Fig. 2(c). The field received at the output and the target fundamental TE mode profiles are plotted in Fig. 2(d). Using a mode expansion monitor, the 3D-FDTD simulation concludes that 99.7\% of the input power is coupled to the fundamental TE mode at the 9-$\upmu$m-wide output of the device. This result verifies the operation of the optimized device, as the vast majority of power is transmitted to the desired output mode at the end of the taper. At the end of the optimization, the only higher-order mode that carries the most non-negligible power is mode \#3, characterized by the three central lobes in the field profile. This accounts for the majority of the 0.3\% of power that is not coupled to the fundamental TE mode.

Additionally, even though our design procedure only considers eigenmode overlaps at a single wavelength ($\lambda$=1550 nm), the resulting taper still achieves a wide optical operating bandwidth. The final 1 dB bandwidth remains above 250 nm, as demonstrated by the 3D-FDTD result in Fig. 2(e). This operation bandwidth exceeds that of many of the highest-performance designs reported to-date [21, 22, 23, 24], while achieving orders of magnitude faster design times and power coupling efficiencies (over 99\%) consistent with these literature results. As an added benefit our approach also allows optimizing for coupling to other spatially symmetric higher order modes, for building related devices such as spatial mode converters without sacrificing computational efficiency. These results illustrate the capabilities of ROMEO in designing high-performance, low-loss, broadband, and fabrication-compatible taper structures with user-specified input and output widths within computational time scales that are thousands of times faster than comparable methods.

\begin{figure*}[!]
\centering
\includegraphics[width=7in]{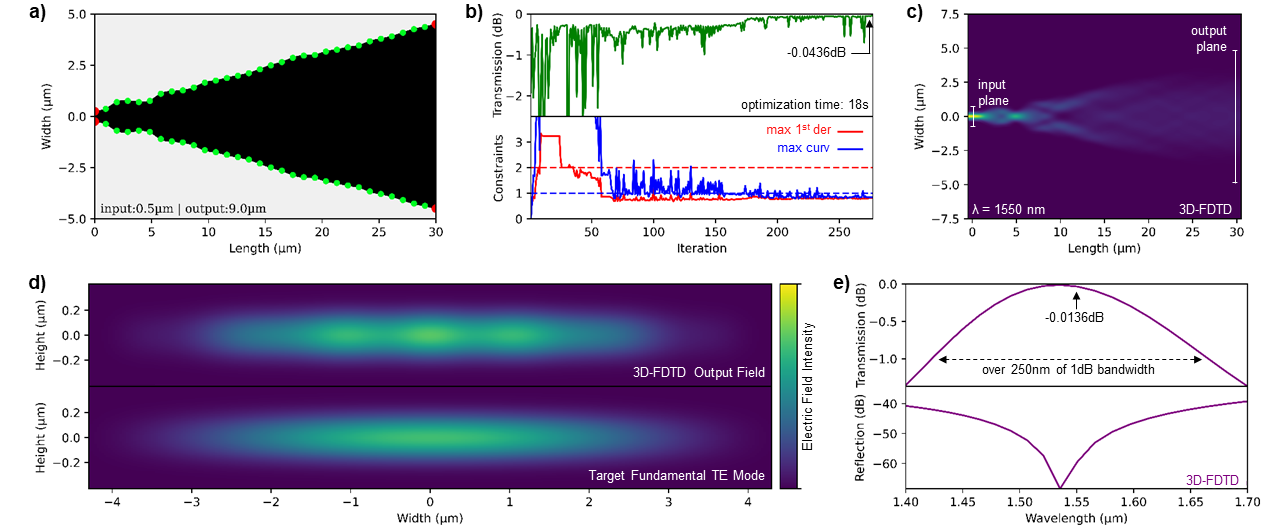}%
\label{fig_first_case}
\caption{Design of waveguide taper from 500 nm to 9 $\upmu$m. a) Optimized device geometry with 30 individually adjustable widths, for a device length of 30 $\upmu$m. b) Optimization progress with transmission to the fundamental TE output mode and constraints. c) 3D-FDTD result of the optimized device. d) The resulting field from 3D-FDTD spatially matches the target fundamental TE output mode, with 99.7\% of the input power coupled to this mode. e) Optical spectrum with -0.0136dB transmission at $\lambda$=1550 nm, a 1dB bandwidth of over 250 nm, and a reflection below -40dB for the entire simulated spectrum.}
\label{fig_2}
\end{figure*}
\section{Design of a 1×2 Power Splitter}
Our database of eigenmode overlaps and the ROMEO framework can be easily adapted for the design of devices with multiple outputs, such as power splitters commonly used in photonic systems. To demonstrate this capability, we design a compact 3 dB splitter constructed from a laterally symmetric, continuous strip waveguide represented by an array of five optimizable widths inside a 2x2$\upmu$m$^2$ footprint. In addition to these optimizable parameters, the input and output waveguides are chosen to be 0.5-$\upmu$m-wide, and the two output waveguides are separated by a fixed gap of 0.2$\upmu$m. As before, the initial guess for this device is chosen using linearly increasing widths between the input and the output as indicated in Fig. 3(a), followed by a 0.2-$\upmu$m-wide fixed separation at the output.

In contrast to the taper example above, this 3 dB coupler requires an additional scattering interface at its output. This interface describes the transmission and reflection of modes as the light propagates between the single-waveguide section of the device through the two-waveguide output section, as shown in Fig. 3(b). Since the output waveguide widths and spacing are fixed, the corresponding scattering matrix for this interface also remains constant throughout optimization. Therefore, it is sufficient to calculate the scattering matrix at this location only once and repeatedly reuse it for the entire device design procedure.

Similar to the taper we demonstrated above, the objective for this device is also configured to maximize the transmission between the fundamental TE modes at the input and output planes. Due to the symmetry of the output waveguides, this specific objective achieves even and in-phase optical power splitting for this 1x2 geometry. As before, the device optimization is performed iteratively with the same optimization framework in ROMEO, using first derivative ($dw(z)⁄dz<2$) and curvature ($\kappa < 4 um^{-1}$) constraints and the COBYLA algorithm. The resulting splitter with nonlinearly optimized widths is illustrated in Fig. 3(b). The optimization procedure for this 3 dB splitter is completed in 445 iterations within a total of 14 seconds, as shown in Fig. 3(c). Similar to the taper design in the previous section, the transmission objective gradually approaches the 100\% target, as the device widths are iteratively modified in search of the optimal device geometry. A final transmission of 97.1\% (-0.1278dB) is achieved between the specified fundamental input and output modes. Here, the transmission objective as well as the two constraints also experience fluctuations due to the exploratory behavior of the algorithm during optimization. At the end, first derivative and curvature constraints are satisfied, indicating a smooth device with fabrication-compatible geometrical features. 

The 3D-FDTD response of this final device showing the resulting electric field at $\lambda$=1550nm and the spectral transmission are plotted in Fig. 3(d) and Fig. 3(e), respectively. Here, we have also added S-bends at the end of the splitter geometry in order to separate the outputs into two distinct and uncoupled waveguides. The additional loss remains below 0.2dB for over a bandwidth of approximately 150 nm, indicating the optimized structure's high performance and broadband operation capability. These 3D-FDTD transmission results closely match our eigenmode expansion calculations, indicating the high physical accuracy of our optimization result. Comparatively, our 1-dB bandwidth of over 300 nm exceeds that of many other compact designs demonstrated [24, 25, 26, 27, 28, 29, 30, 31]. The simulated excess insertion loss of -0.1355dB is consistent with these existing state-of-the-art simulation results from similar 1x2 power splitters in literature. In addition to the resulting device designs’ high performance and compact footprint, our database-driven eigenmode expansion method remains applicable to other interferometers, including 2x2 couplers or 2x4 (90° hybrid) structures, demonstrating its wide operational utility.

It may be possible to further optimize the transmission in this device by modifying the geometry of this two-waveguide output section. This would involve creating another database of effective indices and scattering matrices between modes in two-waveguide structures. Assuming symmetric output waveguides, this new database would need to be constructed as a function of both the waveguide width and the separation. Subsequently, a nonlinear optimizer can be employed to optimize a structure with both single-waveguide and two-waveguide sections in a similar manner. However, building a multi-variate database may be computationally prohibitive, and must be carefully evaluated in comparison to methods based on time-domain or frequency-domain inverse design [32, 33].
\begin{figure*}[!t]
\centering
\includegraphics[width=7in]{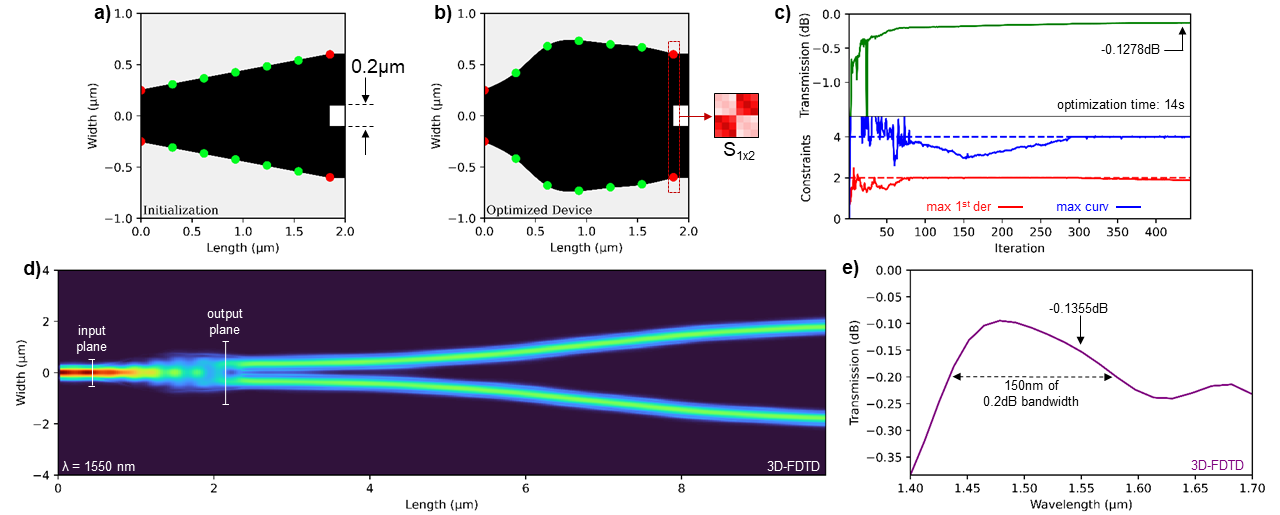}%
\label{fig_first_case}
\caption{Design of 1×2 Splitter. a) Initial device geometry with a 0.2$\upmu$m fixed gap between the outputs. b) Optimized device geometry, with the additional scattering matrix at the output. c) Optimization progress with transmission to the fundamental TE output mode and constraints. d) Electric field intensity from 3D-FDTD simulation of the optimized device, showing even power splitting to the fundamental TE mode. e) Optical spectrum with -0.1355dB transmission at $\lambda$=1550 nm, and a 0.2dB bandwidth of over 150 nm.}
\label{fig_3}
\end{figure*}
\section{Design of a Waveguide Crossing}
Waveguide crossings are one of the critical components of large-scale photonic systems, as they allow optical signals to intersect and cross each other in planar circuit geometries. In this section, we demonstrate the capability of the ROMEO framework to rapidly design on-chip waveguide crossings with very low loss and crosstalk, as well as wide operational bandwidth. Particularly, the 4-fold center-symmetric design of the crossing geometry allows us to represent the trainable widths of the structure using only one of the input waveguides. The initial guess is again specified using linearly increasing widths as shown in Fig. 4(a). Rounded corners are used at the intersections of each input to prevent the formation of sharp features for fabrication compatibility.

For this crossing, in addition to utilizing the ROMEO framework for rapidly obtaining the device response, we further simplify the overall eigenmode expansion procedure by utilizing the underlying device symmetry. Specifically, we first obtain the scattering matrix for the first half of the device through the parallel star products illustrated in Fig. 1(d). We then use the same mathematical procedure described earlier in Section II to create the scattering matrix corresponding to the second (mirrored) half of the device and compute the overall scattering matrix using the star product of the two. Using this device response, we define our objective function in order to maximize the transmission between the fundamental TE modes at the input and output planes, which inherently also minimizes back reflection and crosstalk at other output ports. We use the same first derivative and curvature constraints as before ($dw(z)⁄dz<2$ and  $\kappa < 4 um^{-1}$) and optimize the device until a relative objective convergence of $10^{-5}$ is reached. 

Fig. 4(b) shows the final optimized device structure; and the optimization progress is plotted in Fig. 4(c). The final structure consists of nonlinearly optimized widths, similar to the previous taper and splitter structures we demonstrated. As before, the transmission objective is gradually optimized towards the ideal 0dB target by modifying the specified widths. A final transmission of -0.0436dB is achieved within only 49 iterations. The entire optimization procedure for this device takes only 5 seconds. The maximum derivative and curvature parameters also remain below their thresholds throughout the entire design process, as illustrated by the red and blue curves in Fig. 4(c) respectively. These results indicate that the device remains smooth and fabrication-compatible throughout the entire design process.

Next, we verify the performance metrics of the optimized crossing by running 3D-FDTD simulations on the final geometry. In Fig. 4(d), we plot the resulting electric field intensity at $\lambda$=1550 nm using the fundamental TE-mode input. As specified by our device objective, no obvious coupling to the undesired output ports is observed, and the output field matches well with the desired fundamental mode. We plot the wavelength dependence of the device transmission in Fig. 4(e), indicating a broad functional operation range with 0.1 dB bandwidth reaching as wide as 130 nm and a 1 dB bandwidth of over 300 nm. The reflection and crosstalk obtained from 3D-FDTD results are plotted in Fig. 4(e). This optimal device achieves better than -50 dB reflection and crosstalk throughout the C-band. Over the entire 300 nm simulated bandwidth, both the reflection and crosstalk remain below -43 dB, pointing to the wide operation bandwidth of the resulting optimized device. Even though the overall goal for the optimizer is to maximize the forward transmission at the output plane, this is inherently equivalent to minimizing power lost to any other modes including those traveling in the backward direction as well any higher order modes. Practically, this is how the optimizer indirectly minimizes metrics including reflection and crosstalk, while directly maximizing transmission.

While devices with much narrower operation windows (such as just the C-band) have been shown with losses below 0.01dB [34, 35], achieving low-loss operation with ultra-wideband responses remains an important challenge for crossings. Compared to previously reported results of waveguide crossings in SOI platforms [27, 36, 37], our device demonstrates comparable transmission efficiencies while achieving much wider simulated bandwidths. Moreover, our computationally efficient proof-of-concept demonstrations using ROMEO remain generalizable to other waveguide geometries and material platforms, where these performance metrics can be improved using designs with ridge waveguides or subwavelength structures [38].

\begin{figure*}
\centering
\includegraphics[width=7in]{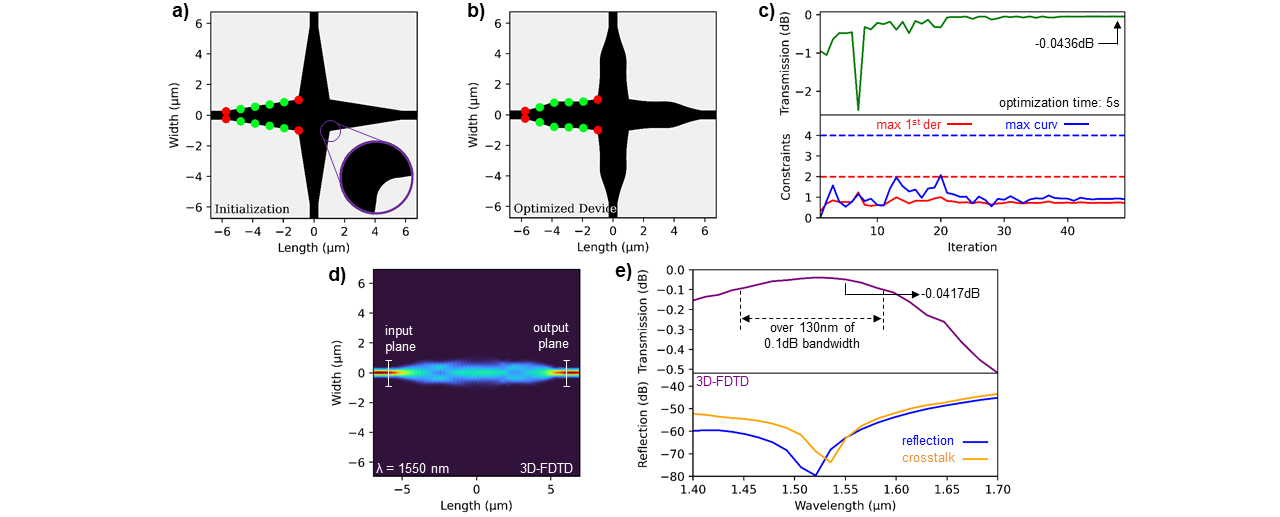}%
\label{fig_first_case}
\caption{Design of waveguide crossing. a) Initial device geometry with rounded corners at waveguide intersections. b) Optimized device geometry. c) Optimization progress with transmission to the fundamental TE output mode and geometrical device constraints. d) Electric field intensity from 3D-FDTD simulations. e) Optical spectrum with -0.0417dB transmission at $\lambda$=1550 nm, a 0.1dB bandwidth of 130 nm, and reflection and crosstalk results below -45dB and -43dB, respectively.}
\label{fig_4}
\end{figure*}

As with any optimization problem, there is a possibility that the optimized device's final performance does not meet the target application’s requirements. In this case, for any one of the three classes of devices we demonstrated here, hyperparameter tuning such as increasing the number of controllable widths might improve the final result, if the device is sufficiently long. However, this is not always guaranteed, as previous studies have shown that there may not be a direct correlation between the optimization outcome and the number of trainable widths [29]. To address this, experimenting with various device lengths, the number and boundaries of controllable widths, thresholds for geometrical constraints, and even different optimization algorithms can help achieve the ideal device for specific applications. ROMEO's underlying computational efficiency makes it practical to perform these experiments and choose the appropriate parameters for application-specific requirements.

\section{Conclusion}
Our rapid eigenmode expansion method represents a significant advancement in data-driven, computationally efficient device design for silicon photonics compared to conventional techniques. This transformative approach enables simulation speeds over 100000 times faster than traditional methods while maintaining high physical accuracy in a flexible design space.

\section*{Acknowledgments}
This work was supported by the Outstanding Young Scientists Awards (GEBİP) program through the Turkish Academy of Sciences. The authors thank Aytug Aydogan for his contributions to the optical data retrieval of this work.

{\appendices
\section*{A. Redheffer Star Product}
Redheffer Star Product operation yields the combined scattering matrix of two consecutive linear scatterers $S^A = 
\begin{bmatrix}
S_{11}^{A} & S_{12}^{A} \\
S_{21}^{A} & S_{22}^{A} 
\end{bmatrix}$ and $ S^B = 
\begin{bmatrix}
S_{11}^{B} & S_{12}^{B} \\
S_{21}^{B} & S_{22}^{B} 
\end{bmatrix} $ as $S^{AB} = S^AS^B =  \begin{bmatrix}
S_{11}^{AB} & S_{12}^{AB} \\
S_{21}^{AB} & S_{22}^{AB} 
\end{bmatrix} $  where the individual blocks of the  star product are calculated through

\begin{equation} 
\label{redheffer}
\begin{aligned} 
S_{11}^{AB}&= S_{11}^{A} + S_{12}^{A}[I-S_{11}^{B}S_{22}^{A}]^{-1}S_{11}^{B}S_{21}^{A} \\
S_{12}^{AB}&= S_{12}^{A}[I-S_{11}^{B}S_{22}^{A}]^{-1}S_{12}^{B} \\
S_{21}^{AB}&= S_{21}^{B}[I-S_{22}^{A}S_{11}^{B}]^{-1}S_{21}^{A} \\
S_{22}^{AB}&= S_{22}^{B} + S_{21}^{B}[I-S_{22}^{A}S_{11}^{B}]^{-1}S_{22}^{A}S_{12}^{B} 
\end{aligned}
\end{equation}

\section*{B. Optimizer Convergence under Different Initial Conditions}

The initial conditions can strongly influence the convergence speed and the performance of the final device. To explore these, we have performed 15 optimizations with random initial conditions for all three of our devices, up to a maximum of 200 iterations as plotted in Fig. 5.

\begin{figure}[!h]
\centering
\includegraphics[width=3.5in]{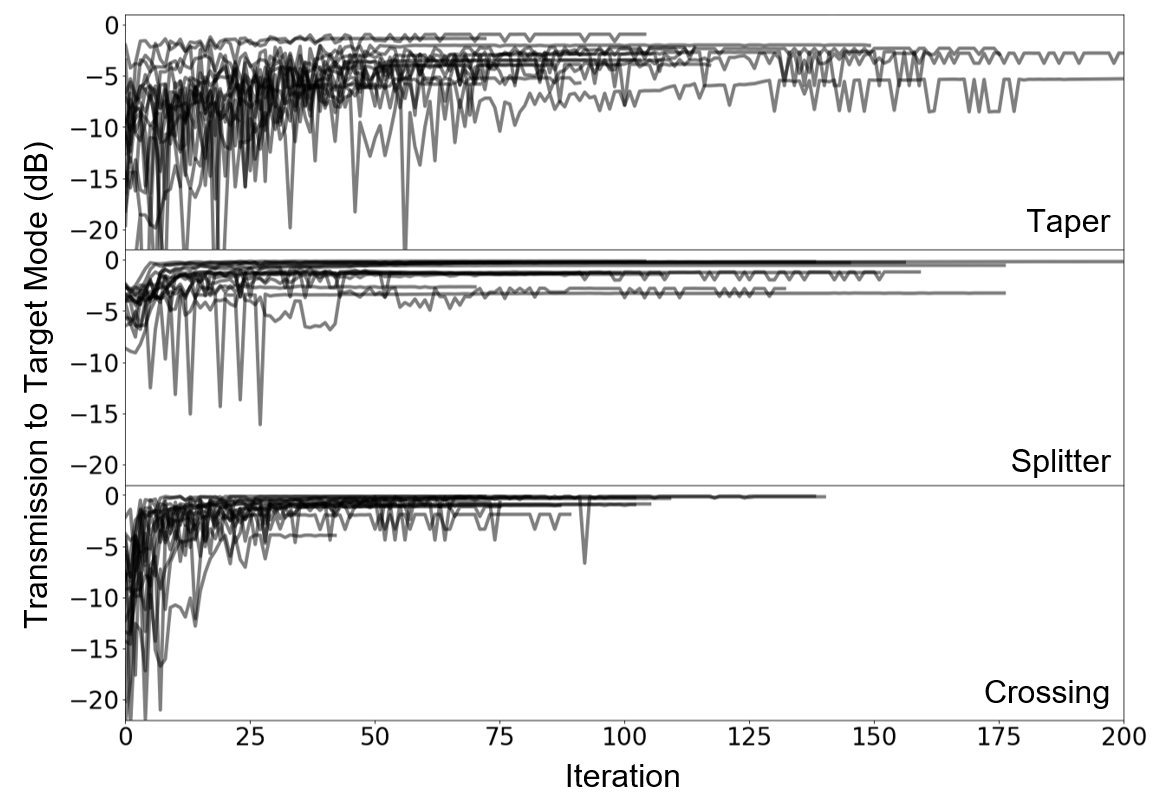}
\caption{Convergence of the three devices under different initial conditions.}
\label{fig_5}
\end{figure}

Our results indicate that most devices show significant improvement in their objective function, irrespective of their initial starting conditions. Despite these improvements, some optimizations remain confined to local extrema and do not yield acceptable final performance metrics, as their objective functions plateau after the first 50-100 iterations. For all devices, the time to reach 100 iterations typically remains below 10 seconds, after which the convergence behavior of the device can easily be evaluated by the designer. This performance is enabled by the computational efficiency of the individual eigenmode simulations, as we demonstrated in Fig 1(d). However, if a particularly unfavorable initial condition is specified, the target objective may never be reached. This occurrence is related to the optimizer's ability to effectively search the multi-dimensional objective space starting from a poor initial condition, and is regardless of the computational efficiency of the underlying simulations. This is not unique to optimization using EME or photonic design, but is actually a well-documented phenomenon in machine learning [39, 40]. To mitigate these problems, it is common practice to employ “multi-start” optimizations to explore different regions of the solution space more thoroughly, and improve the likelihood of an acceptable solution. Additionally, if a physically motivated guess or an approximate solution exists (similar to those in Fig 3(a) and 4(a)), it can serve as a beneficial starting point in the search for the ideal solution.}

\newpage

\end{document}